\documentclass[aps,pra,a4paper,superscriptaddress,longbibliography,reprint]{revtex4-1}

\makeatletter
\def\@fnsymbol#1{\ensuremath{\ifcase#1\or *\or * \else\@ctrerr\fi}}
\makeatother

\usepackage[T1]{fontenc}
\usepackage{newtxmath,newtxtext}

\usepackage{amsmath}
\usepackage{amsfonts}
\usepackage{bm}
\usepackage{bbm}

\usepackage{graphicx}

\usepackage{color}

\newcommand{\eq}[1]{Eq.~(\ref{#1})}
\newcommand{\fig}[1]{Fig.~\ref{#1}}

\newcommand{\dKS}{$\Delta$KS}
\newcommand{\etal}{\textit{et al.}}

\begin{document}

\title{Accurate computational prediction of core-electron binding energies in carbon-based
materials: A machine-learning model combining density-functional theory and $\boldsymbol{GW}$}

\author{Dorothea Golze}
\email{dorothea.golze@tu-dresden.de}
\affiliation{Faculty of Chemistry and Food Chemistry, Technische Universit\"at Dresden, 01062 Dresden, Germany}
\affiliation{Department of Applied Physics,
Aalto University, 02150, Espoo, Finland}

\author{Markus Hirvensalo}
\affiliation{Department of Applied Physics,
Aalto University, 02150, Espoo, Finland}

\author{Patricia Hern\'andez-Le\'on}
\affiliation{Department of Electrical Engineering and Automation,
Aalto University, 02150, Espoo, Finland}

\author{Anja Aarva}
\affiliation{Department of Electrical Engineering and Automation,
Aalto University, 02150, Espoo, Finland}

\author{Jarkko Etula}
\affiliation{Department of Chemistry and Materials Science,
Aalto University, 02150, Espoo, Finland}

\author{Toma Susi}
\affiliation{University of Vienna, Faculty of Physics, Boltzmanngasse 5, 1090, Vienna, Austria}

\author{Patrick Rinke}
\affiliation{Department of Applied Physics,
Aalto University, 02150, Espoo, Finland}

\author{Tomi Laurila}
\affiliation{Department of Electrical Engineering and Automation,
Aalto University, 02150, Espoo, Finland}
\affiliation{Department of Chemistry and Materials Science,
Aalto University, 02150, Espoo, Finland}

\author{Miguel A. Caro}
\email{mcaroba@gmail.com}
\affiliation{Department of Electrical Engineering and Automation,
Aalto University, 02150, Espoo, Finland}

\date{\today}

\begin{abstract}
\noindent \textbf{Abstract}
\\[1em]
We present a quantitatively accurate machine-learning (ML) model for the computational prediction
of core-electron binding energies, from which x-ray photoelectron spectroscopy (XPS) spectra
can be readily obtained. Our model combines density
functional theory (DFT) with $GW$ and uses kernel ridge regression for the ML predictions.
We apply the new approach to disordered materials and small molecules containing
carbon, hydrogen and oxygen,
and obtain qualitative and quantitative agreement with experiment, resolving spectral
features within 0.1~eV of reference experimental spectra. The method
only requires the user to provide a structural model for the material
under study to obtain an XPS prediction within seconds. Our new tool is freely
available online through the XPS Prediction Server.
\end{abstract}

\maketitle

\section{Introduction}

X-ray spectroscopy techniques are routinely used for structural characterization of 
materials~\cite{Bagus2013} and molecules~\cite{Siegbahn1969all}. Among the different
x-ray spectroscopies, x-ray photoelectron
spectroscopy (XPS) is arguably the most widespread. In XPS, a material sample is irradiated
with monochromatic x-rays to probe the binding energy (BE) of its core electrons.
When a core electron absorbs an x-ray photon with
enough energy, it leaves the sample and its kinetic energy can be measured. Since the x-ray incident
energy is known, the difference between the kinetic and the incident energy is the core-electron BE. This
energy is characteristic of the chemical environment of the core-excited atom. XPS spectra are
therefore frequently used for structural characterization~\cite{Bagus1999,Siegbahn1969all}.

Carbon-based materials, such as amorphous and disordered carbons, graphene, graphene derivatives,
and nanotubes are an important materials class in industry
and research~\cite{robertson_2002,geim_2007,savazzi_2018}. Furthermore, 
emerging applications are envisioned, such as energy storage and
conversion~\cite{vix_2005}, electronics~\cite{santini_2015},
electrocatalysis~\cite{tavakkoli_2015,cilpa_2019} and biosensing~\cite{laurila_2017}.
Unfortunately, 
the atomic structure of carbon-based materials is often not completely known because,
in addition to their possibly disordered nature, they also often contain a wide variety
of defects and surface chemical functionalizations.
XPS is one of the most commonly used spectroscopy tools for structural characterization of
carbon-based materials~\cite{sainio_2016,laurila_2017}. However, it is generally difficult, and
sometimes impossible, to establish the precise origin of each peak in an XPS spectrum due to the
lack of well-defined reference data~\cite{aarva_2019}. In addition, the link between atomic
structure and a particular XPS spectrum is often imprecise since the core-electron BEs of two atoms
in different
chemical environments can be the same. Further ambiguities are introduced by the peak fitting procedure,
which must be applied to resolve overlapping features in the experimental spectrum and which often
relies on a number of assumptions, such as the total number of peaks~\cite{aarva_2019,aarva_2019b}.
These limitations impact the interpretation of experimental XPS spectra. Inferring the atomic structure
from the XPS spectrum is referred to as the ``backward'' route.
An alternative strategy is to generate XPS spectra from candidate structural models.
The best match between generated and measured spectra then provides the best structural model.
We call this the ``forward'' direction. 

For the forward route to be feasible, we require a pool of candidate structures and a theoretical
or computational approach that is able to accurately predict core-electron BEs from the atomic
structure alone. Candidate structures can be generated computationally.
The difficulty with the forward direction therefore lies with the  availability and computational
cost of accurate XPS prediction tools. 
Currently, XPS modeling is almost exclusively based on
Kohn-Sham density functional theory (DFT) employing so-called
$\Delta$-methods, such as the Delta self-consistent
field ($\Delta$SCF)~\cite{Bagus1965} or Delta Kohn-Sham ($\Delta$KS) frameworks~\cite{ljungberg_2011}.
DFT is, by design, a ground-state theory and does not provide systematic access to excited-state properties.
However, for small molecules, relative and absolute core-level BEs from $\Delta$SCF and $\Delta$KS
generally compare well to experiment, in particular when employing meta-generalized gradient approximation
(meta-GGA) functionals~\cite{Bellafont2016,Bellafont2016b,Kahk2019,Hait2020}. Recently, $\Delta$SCF
calculations with meta-GGAs were, in combination with finite-size correction schemes, also successfully
applied to simple solids~\cite{kahk_2021}.

While the functional dependence of molecular core-level excitations is moderate for absolute~\cite{Bellafont2015,Klein2021} and even
negligible for relative BEs~\cite{Bellafont2016}, it can be more severe for complex
materials~\cite{susi_2015,Susi2018,Hall2021}. As a consequence of the self-interaction error, the accuracy of
DFT-based $\Delta$-methods deteriorates with increasing system sizes, which has been comprehensively
discussed for valence excitations~\cite{Pinheiro2015} and has been also observed for core states~\cite{Golze2018}.
The $GW$ approximation~\cite{Hedin1965} to many-body perturbation theory overcomes these limitations of DFT
and provides a rigorous quantum-mechanical framework for the photoemission process~\cite{Golze2019},
see also Refs.~\cite{Golze2018} and \cite{Golze2020} for a discussion of the limitations of $\Delta$-methods. However,
the computational cost of a $GW$ calculation is orders of magnitude larger than for DFT, making a
straightforward application of $GW$ to complex materials difficult.
 
The computational prediction of XPS spectra of amorphous structures requires
sampling over all atoms in the structure, which is demanding even with DFT,
and impossible with $GW$. Machine-learning (ML) methods are a promising strategy to bridge
the gap between high accuracy and computational efficiency. While the application of ML in
spectroscopy is still in its infancy~\cite{Stuke/etal:2019, Gosh/etal:2019,Westermayr/Marquetand:2020,
Dral/Barbatti:2021}, the first proof-of-concept applications for the prediction of valence~\cite{Westermayr2021}
and core-level spectra
are emerging~\cite{Carbone2019,Carbone2020,Lueder2021}. In this paper, we advance these ideas for
real-world applications and develop a powerful XPS prediction tool by combining $\Delta$KS calculations and
highly accurate $GW$ predictions with ML models. We show that our method can provide access to quantitatively
accurate predictions of XPS spectra of complex disordered materials and small molecules containing carbon, hydrogen and oxygen
(CHO).

The remainder of this article is structured as follows: We discuss the generation of structural models for
CHO-containing compounds in Section~\ref{sec:structuralmodels}. We proceed with the details of the
electronic-structure methods used to generate computational XPS data in Section~\ref{sec:dftandgw}. The
architecture of our ML models is discussed in Section~\ref{sec:MLmodel}. The performance of the ML models for
C${1s}$ and O${1s}$ excitations of molecular and extended CHO structures is presented in Section~\ref{sec:results},
followed by XPS spectra predictions for selected CHO materials. Finally, we introduce our XPS Prediction Server as
a freely available online tool and draw conclusions in Section~\ref{sec:summary}.

\section{Methods}

\subsection{Structural models of CHO-containing compounds}
\label{sec:structuralmodels}
Due to carbon's versatility in chemical bond formation, the composition and configuration space for
carbon-based materials and molecules is vast~\cite{kirkpatrick_2004,vonlilienfeld_2020}.
Diverse examples of CHO compounds are small molecules
(water, methane, methanol, etc.), large molecules and polymers (lipids, sugars, cellulose, etc.),
and solid-state materials. The main focus of this work are CHO materials. While two distinct CHO
materials will differ from each other when regarded
as a whole, they are made of the same (or very similar) individual building blocks, or atomic
motifs. By building a library of structural models for CHO materials, we can identify the individual atomic motifs most
representative of the ensemble~\cite{caro_2018c}. Effective motif selection, discussed in more detail
in Section~\ref{sec:classification}, is essential to obtain a compact and manageable representation of large structural
databases.

\subsubsection{CHO structural databases}

We have generated two CHO  structure databases for the prediction of XPS data: one for CHO materials
and another one for small CHO-containing molecules. The structural database for CHO materials was constructed from
computer-generated model structures of amorphous carbon (a-C), hydrogenated a-C (a-C:H), oxygen-enriched
a-C (a-C:O), functionalized a-C, oxygenated amorphous carbon (a-CO$_x$), graphene (G), and reduced
graphene oxide (rGO). All of the computational structural models for these materials are taken from the available
literature~\cite{kumar_2013,caro_2014,susi_2014,deringer_2017,scardamaglia_2017,caro_2018c,deringer_2018}.
An exception are the oxygen-rich a-CO$_x$ models, which were generated using DFT molecular dynamics following Ref.~\cite{deringer_2018}.
The CHO materials in our structural database cover a broad range of structural building blocks, which provides
the necessary foundation to
map all the characteristic
atomic motifs centered on O and C (H lacks an atomic core) to their corresponding core-electron binding
energies.

Our structural database of small CHO molecules is a subset of the QM9 data set~\cite{ramakrishnan_2014},
which contains in total 134k organic compounds. Our QM9 subset consists of 2089 CHO molecules with 3--29
atoms and its size distribution follows that of those QM9 molecules which contain exclusively C, H and O.
The full QM9 database also includes molecules with N and F. Those molecules were not considered for the
subset generation. The molecules in our subset contain up to 9 ``heavy'' atoms (in this case C or O), which
amounts to a total of 14,707 C$1s$ and 1865 O$1s$ excitations. We found that the relationship between local
molecular structure and XPS properties is not transferable to extended CHO structures, i.e., these data were
not used for the generation of the ML models for CHO materials.  However, since these molecules are small,
calculating their XPS spectra is computationally comparatively inexpensive. This allows us to both
benchmark our methodology in the limit of abundant data and produce a useful reference database of
computational XPS spectra of small CHO-containing molecules, which is valuable on its own.

\subsubsection{Structure classification based on representative atomic motifs}
\label{sec:classification}
The computational structures for the CHO materials contain many C and O sites. Computing the $1s$
core-electron BE for each site is, unlike for molecules, computationally too expensive, in particular at
the $GW$ level of theory.
Therefore, to make the
computational effort tractable, we have identified the most representative atomic motifs in
the entire CHO materials database, using the data clustering methodology presented in
Ref.~\cite{caro_2018c}. In short, a many-body atomic descriptor known as the
``smooth overlap of atomic positions'' (SOAP) is used to encode the atomic structure
surrounding each atomic site in the database of structures~\cite{bartok_2013}. These descriptors
allow us to construct kernel functions, which can also be understood as measures of
similarity between the atomic environments. From these, we can build a distance matrix
that can then be fed to a data classification algorithm. This algorithm, $k$-medoids
in our case~\cite{bauckhage_2015}, clusters data (here the atomic environments) into groups
that share similarities, and assigns a ``medoid''
(also called a ``centroid'' in barycenter-based clustering methods). This medoid corresponds
to the most representative atomic motif within each data cluster. We can preselect the number
of data clusters to build, based on our estimate of available CPU power, and perform the
core-electron calculations only on those. This leads to efficient charting of configuration
space, since we avoid repeatedly calculating motifs that are overrepresented in the database of
atomic structures.

\begin{figure}
\centering
\includegraphics[width=\columnwidth]{{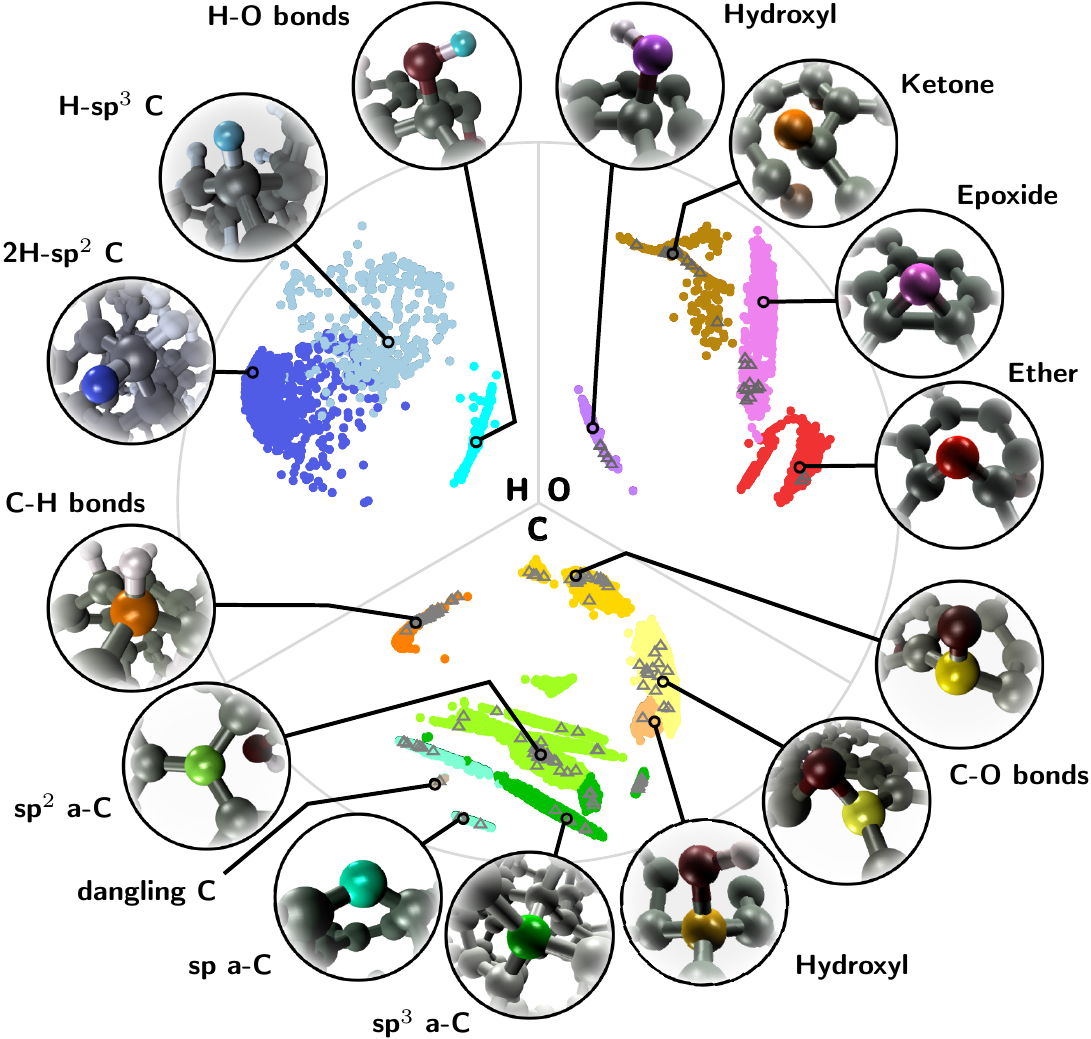}}
\caption{Cluster-based multidimensional scaling map of the CHO materials database used in this
work. The graph is partitioned into three sections, depending on whether a C, H or O atom
is at the center of the environment. For instance, a hydroxyl group is viewed differently
depending on whether the atomic environment descriptor uses C, H or O as the origin.
The distance between data points on the map is inversely proportional to the
degree of similarity between the corresponding atomic environments. This similarity
is established using SOAP many-body descriptors~\cite{bartok_2013,caro_2019},
as explained in more detail in Refs.~\cite{de_2016,cheng_2020,caro_2018c}. The gray triangles
indicate the motifs selected for the $GW$ calculations.}
\label{16}
\end{figure}

The composition of the database is visualized in \fig{16}, where we show
a map of chemical and structural similarities between the present atomic motifs using
a low-dimensional embedding tool, cl-MDS~\cite{hernandez-leon_2021}, which combines ML atomic
descriptors~\cite{bartok_2013, caro_2019} and multidimensional
scaling (a dimensionality reduction technique)~\cite{de_2016,caro_2018c,cheng_2020}
with data clustering~\cite{bauckhage_2015}.
Here, an atomic motif is constructed from a central C, H or O atom embedded within diverse
CHO environments. Atomic motifs are associated in classical terms with coordination
environments (e.g., $sp$, $sp^2$ and $sp^3$ in carbon~\cite{caro_2018c}) or with chemical groups,
such as keto, epoxide, hydroxyl groups, etc. The central atom is either part of this group or adjacent to
it, see \fig{16}. The environment within the immediate vicinity of the central atom includes all
the neighbor atoms within a given cutoff sphere. To simplify visualization, we used a cutoff
radius of $\sim 2.25$~\AA{} to generate the similarities in \fig{16}. However, this radius is
too small when selecting structures for training the ML models. Motivated by the convergence
studies for finite systems discussed in Section~\ref{sec:carving}, we retain structural
information within a cutoff sphere of radius 4.25~\AA{} for all the ML models trained in this
study.

The subset of atomic motifs in the database selected
for $GW$ calculations is indicated with gray triangles in \fig{16}. \dKS{} calculations are
performed on an extended subset of the whole database that includes the $GW$ environments.
Note that core-level calculations are only performed for heavy atoms and we include thus
only motifs with central C or O atom. Using the structures selected for $GW$ as an example
(the gray triangles), \fig{16}
visually highlights two aspects of motif selection: i) The diversity of our CHO database is
preserved in the selection of those structures used for ML model construction, i.e., we draw samples
from all over the map. ii) This selection needs to be carried out according to the longer cutoff,
since a classification based on a shorter cutoff, while useful in visualization, does not contain
enough information to train a predictive ML model (the drawn samples are not homogeneously
distributed on the map). If we
drew the same map according to similarities based on the 4.25~\AA{} cutoff, \fig{16} would look
very different and it would be impossible to make an intuitive connection between the map
and classical chemical group classification. See Fig.~S5 of the Supporting Information (SI)
for a cl-MDS map with the larger cutoff.

\subsubsection{Carving out the structures}
\label{sec:carving}
\begin{figure}
\centering
\includegraphics[width=1\columnwidth]{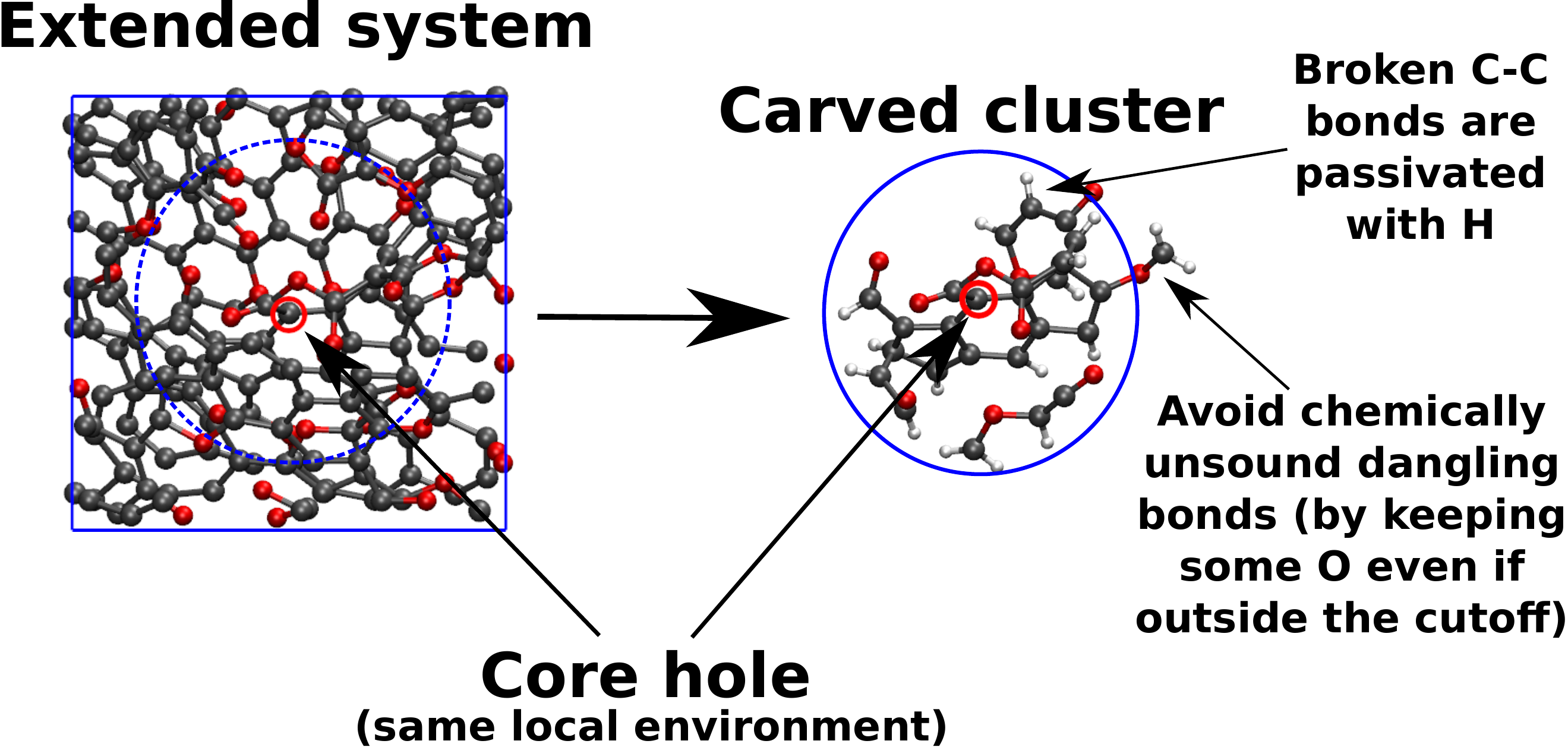}
\\[0.5em]
\includegraphics[width=1\columnwidth]{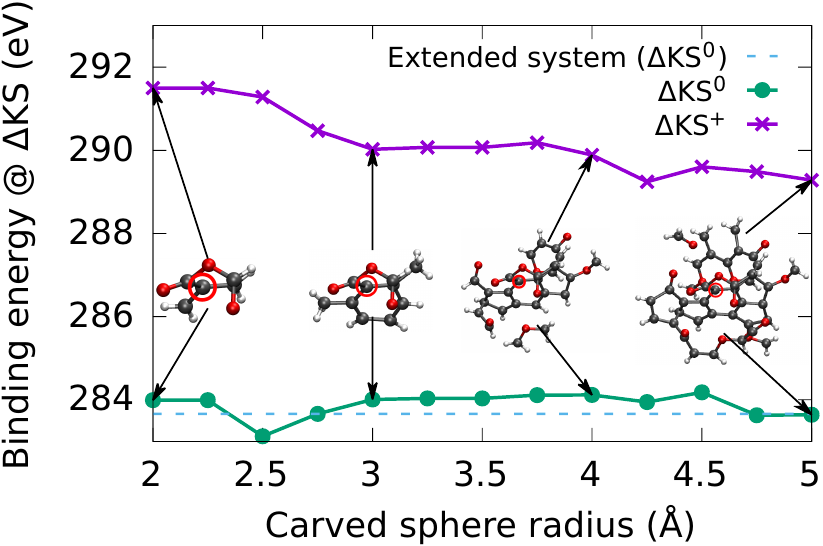}
\\[0.em]
\includegraphics[width=1\columnwidth]{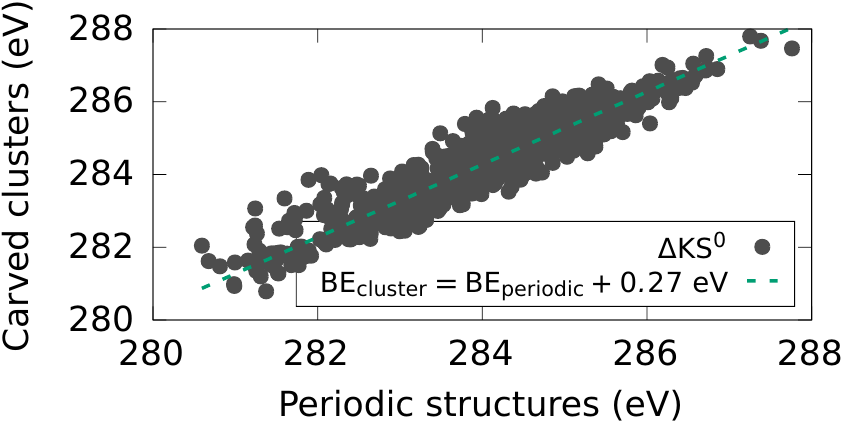}
\vspace{-2em}
\caption{(Top) Example of the cluster carving procedure for an a-CO$_x$ structure,
where the cluster is contained within a sphere centered on the
carbon atom highlighted with the red circle where the core hole is created.
(Middle) Dependence of the core-electron BE of the cluster on the cutoff radius, together with the periodic reference,
for the example structure in the top panel ($\Delta$KS values).
(Bottom) Comparison for the whole set of $\Delta$KS data points in our database
for which
both cluster ($r_\text{cut} = 4.25$~{\AA}) and periodic $\Delta\text{KS}^0$ values are available, showing that the main difference is simply
a small vertical shift in the energies. This strongly indicates that the
carved structures are good surrogate models for the extended (periodic) systems. $\Delta\text{KS}^+$ refers to calculations
where the excited core electron is removed from the system, whereas in $\Delta\text{KS}^0$ it is promoted to the conduction band, 
see Section~\ref{13}. The root-mean-square and mean-absolute errors, once the rigid
shift is taken into account, are 0.41~eV and 0.32~eV, respectively, with a maximum error of
1.69~eV.}
\label{01}
\end{figure}

Another issue pertains to the size of the input structures, which are model ``supercells''
with up to 355 atoms with periodic boundary conditions. The cost of DFT
calculations scales cubically with the number of atoms in the system.  The cost of core-level $GW$
calculations is even more formidable and scales with the fifth power of the number of
atoms~\cite{Golze2018}. We therefore resort to  moderately sized cluster models of around 100 atoms
for our carbon structures.
Cluster models are justified for core-level excitations, since local atomic properties,
such as the core-electron binding energy, are expected to converge with respect to the number of
neighbors explicitly included in the calculation. By
only keeping the atoms in the immediate surrounding of the core-excited atom, we can significantly
speed up the core-level calculations. At the same time, we remove
the need for periodic boundary conditions, which are currently incompatible with our core-level $GW$
implementation.

The cluster models need to be constructed with care, since creation of dangling bonds or radicals
at the cluster surface can affect the overall spin state of the system. We use a ``carving''
technique introduced in Refs.~\cite{caro_2014,caro_2020c}, where a spherical portion of the
material centered on a specific site is carved out of the bulk material. Broken C-C bonds are
passivated with H atoms. All other bonds (C-O, O-H and C-H) are preserved. The procedure is
exemplified in \fig{01} and our carving code is freely available online~\cite{carvef90}.

We find that the core-level BEs converge quickly with respect to the cluster radius for $\Delta$KS,
as shown in \fig{01} (middle), and also for $GW$, see Fig.~S1 in the SI.
Convergence is typically reached for $r_\text{cut} = 4.25$~{\AA}, which is the cluster
radius we use for the data acquisition of our ML models. Figure~\ref{01} (bottom) also indicates
that the carved structures represent a good surrogate model for the periodic structures, since
the respective predicted core-electron binding energies closely follow each other. The main
difference is a constant shift of 0.27~eV, which is easy to correct for. We attribute
the systematic 0.27~eV upward shift to finite-size (or particle-in-a-box) effects upon promotion
of the core electron to the valence. Since the electron localization length is necessarily reduced
in the 4.25~\AA{} cluster compared to the extended structure, the energy of the excited state
increases accordingly, by 0.27~eV on average in this case.
The validity of the cluster approach is further discussed in Section~\ref{sec:MLmodelsCHO}.

\subsection{DFT and \textit{GW} calculations}
\label{sec:dftandgw}
\subsubsection{DFT}\label{13}

We carry out the DFT calculation of core-electron BEs using the \dKS~\cite{ljungberg_2011}
total energy method. \dKS{} is computationally more affordable,
and therefore more amenable to high-throughput calculations,
than its similarly accurate all-electron variant, the
$\Delta$SCF~\cite{Bagus1965} method (see the SI for an explanation regarding the difference between
the $\Delta$KS and $\Delta$SCF method). In $\Delta$KS, the core-level BE is given as the difference between
core-excited and ground-state total energies. In the excited state calculation, the C1$s$ or O1$s$
electron is removed from the core, which is modeled via a special projector augmented-wave
(PAW~\cite{bloechl_1994}) potential, and only the valence electrons are relaxed.
These valence electrons can be relaxed either in the presence of the excited electron
(neutral calculation, \dKS$^0$) or in its absence (charged
calculation, \dKS$^+$). For molecules, the \dKS$^+$ approximation can be applied directly
since the vacuum level is well defined. For materials, the \dKS$^0$ calculation allows to
align the computed BEs similarly as in experiment, i.e., with respect
to the Fermi level. In addition to the periodic \dKS$^0$ calculations, which we need
for our ML model,
we carried out \dKS$^0$ calculations also for carved clusters to validate that
these clusters are indeed good surrogate models for the periodic structures, as shown in
\fig{01} (bottom). However, the \dKS$^+$ values are those directly comparable to
our $GW$ cluster calculations. 

We performed open-shell DFT calculations with VASP~\cite{kresse_1996,kresse_1999,kohler_2004}
using the Perdew-Burke-Ernzerhof (PBE) functional~\cite{perdew_1996}. See the SI for
a discussion on the choice of functional.
We apply a constant correction based on
GPAW~\cite{enkovaara_2010,ljungberg_2011} results, to convert the relative
{\dKS} values (i.e., the chemical shifts) from VASP to absolute {\dKS} values.
Disordered carbon materials often exhibit local atomic
magnetization~\cite{laurila_2017}, which makes the determination of the ground state challenging.
We expand on the procedure how to determine the lowest energy magnetic configuration, details of
the DFT calculations and the definition of the reference level for comparison to experiment in the SI.
Further discussions of energy referencing can also be found in Section~\ref{12} and
Ref.~\cite{aarva_2019}.

\subsubsection{GW}

The $GW$ approximation~\cite{Hedin1965} is a highly accurate electronic-structure
method that can be applied to predict photoemission spectra. The central object of $GW$ is the
self-energy $\Sigma$, which is computed from the Green's function $G$ and the screened
Coulomb interaction $W$, where $\Sigma=\text{i}GW$; hence the name $GW$. The self energy
contains all quantum mechanical correlation and exchange interactions between the
electrons and the hole created upon photoemission. $GW$ offers access to quasiparticle
energies, which directly correspond to the negative of the vertical ionization potentials.

$GW$ has become the gold standard for the computation of band structures of solids and is
now also increasingly applied to molecular valence excitations~\cite{Golze2019}. Recently,
we advanced the $GW$ methodology and implementation for application to deep core excitations
by combining exact numeric algorithms in the real frequency domain~\cite{Golze2018}
with partial self consistency~\cite{Golze2020} and relativistic corrections~\cite{Keller2020}.
We showed that $GW$ reproduces absolute molecular 1$s$ excitations within 0.3~eV of experiment
and relative binding energies with average deviations below 0.2~eV~\cite{Golze2020}. 
Our core-level $GW$ approach was recently also applied to simple solids~\cite{Zhu2021},
yielding first promising results. In addition, its extension to the
Bethe-Salpeter equation (BSE@$GW$) was lately also successfully used for the prediction of molecular $K$-edge transition energies~\cite{Yao2022}.

$GW$ calculations are several orders of magnitude more expensive computationally than
DFT calculations with GGA and even hybrid functionals. Nevertheless, $GW$ is nowadays
routinely applied to predict valence excitations of systems with several hundred
atoms~\cite{Wilhelm2016,Stuke2020,Wilhelm2018,DelBen2019,Wilhelm2021,Duchemin2021}.
However, the application of $GW$ to deep core excitations is computationally more
expensive than for valence states. First, core-level $GW$
calculations require more advanced numerical schemes~\cite{Golze2018}, increasing the conventional
scaling with respect to system size $N$ from $O(N^4)$ (valence states)
to $O(N^5)$ (core states). This unfavorable scaling restricts the accessible system size
in core-level $GW$ to around 100 atoms. Second, an all-electron treatment is necessary,
which we efficiently realize by an implementation with localized basis sets. The implementation
of $GW$ in localized basis set codes is a rather recent development of the last
decade~\cite{Golze2019}, for which the efficient implementation of periodic boundary conditions
is still the subject of ongoing work~\cite{Wilhelm2017,Ren2021,Zhu2021}.
Our core-level $GW$ implementation~\cite{Golze2018} is thus currently restricted to cluster
calculations. The largest $GW$ calculation in this work was performed for an a-C cluster with
112 atoms on more than 8000 CPU cores.

For the $GW$ calculations, we use the FHI-aims program package~\cite{Blum2009,Ren2012}
and follow the procedure developed in Ref.~\cite{Golze2020}. We employ a single-shot $G_0W_0$
approach in combination with the PBEh($\alpha=0.45$) functional for the underlying DFT calculation,
where $\alpha$ is the amount of exact Hartree-Fock exchange~\cite{Atalla2013}. The $\alpha$ value
was tuned to reproduce the results of computationally more demanding eigenvalue-self-consistent $GW$
methods~\cite{Golze2020}. We performed a screening for the lowest-energy configuration at the
PBEh($\alpha$) level to ensure that the open-shell $G_0W_0$ calculations are performed on top of the
DFT ground state. All $GW$ results are extrapolated to the complete basis set limit and relativistic
corrections~\cite{Keller2020} are added for the O1$s$ excitations. Further details are given in the
SI. To support open data-driven materials science~\cite{Himanen2020}, we uploaded the input and
output files of all $GW$ calculations of the a-C clusters to the Novel Materials Discovery (NOMAD)
repository~\cite{nomad_repo}.

\subsection{Machine-learning model}
\label{sec:MLmodel}
We develop ML models for the prediction of either a core-level BE for atom $i$, BE$_i$, or the
difference $\Delta_i$ between $GW$ and DFT predicted core-level BEs,
$\Delta_i = \text{BE}_i^{GW} - \text{BE}_i^{\text{DFT}}$. Since all our ML
models use the same architecture, we denote the
quantity to be learned generically as $\gamma_i$. Separate models are trained for C$1s$ and
O$1s$, which is motivated by the structure of our data: core-level BEs are strongly species dependent
and separated by more than 100~eV for different atomic species. C$1s$ excitations occur around 290~eV,
whereas O$1s$ photoelectrons are ejected at approximately 540~eV. The chemical shifts due to different
local atomic environments around a carbon or oxygen core are 2--3 orders of magnitude smaller.

Our ML model is based on kernel ridge regression (KRR), using kernels constructed from \texttt{soap{\_}turbo} descriptors,
which are a modification~\cite{caro_2019} of the SOAP many-body atomic
descriptor~\cite{bartok_2013}, providing improved speed and accuracy. Here, we used the
Python interface (``Quippy'') to the \texttt{soap{\_}turbo} library provided by the QUIP and GAP
codes~\cite{bartok_2010,libatoms}.
Briefly, KRR replaces the non-linear problem of expressing
$\gamma_i$ for the core of atom $i$ as a function of atomic positions,
$\gamma_i = \gamma (\{\textbf{r}_j \in S_i \})$, where $j$
runs through all atoms within an environment of $i$, $S_i$, with a linear problem.
Using the ``kernel trick'', the same quantity, $\gamma_i$, is expressed as a
linear combination of kernel functions, $k(i,t)$:
\begin{align}
\gamma_i = e_0 + \delta^2 \sum_{t = 1}^{N_\text{t}} \alpha_t k(i,t),
\qquad k(i,t) = (\textbf{q}_i \cdot \textbf{q}_t)^\zeta.
\label{02}
\end{align}

Here, $\textbf{q}_i$ are the SOAP-type many-body atomic descriptors that we use
to encode the atomic information
about the environment of atom $i$, and $t$ denotes a number of reference environments
in the training set. The dot-product SOAP kernel $(\textbf{q}_i \cdot \textbf{q}_t)^\zeta$
(where $\zeta = 2$ in our case) provides a measure of similarity between $i$ and $t$
that is rotationally and translationally invariant~\cite{bartok_2013}. $\delta$ is
a parameter, given in eV, which controls the energy scale, and $e_0$ is a
constant reference
energy subtracted during training and then added during prediction.
Conceptually and methodologically the present approach is similar to that
of the Gaussian approximation potential (GAP) formalism~\cite{bartok_2010,bartok_2015},
and to our
previous models of adsorption energetics in carbon-based materials~\cite{caro_2018c}.
The SOAP descriptors encode atomic structural information up to a certain cutoff
radius from the central atom $i$. Thus, we implicitly make the assumption of locality
for the binding energies. That is, we assume that only the arrangement of atoms in the
immediate vicinity of atom $i$ affects the core levels of that atom. The validity
of this assumption is illustrated in \fig{01} (middle), where we show that the core-level
BE of the central C atom  quickly converges with the cluster size. Further confirmation is
obtained from \fig{01} (bottom), which shows that the difference between the C$1s$ excitation from
periodic and cluster models is mainly a constant shift.

Model training consists essentially in the inversion of \eq{02} (or, more
precisely, on a least-squares based optimization of the $\alpha_t$) for a set of
reference calculations, i.e., during training both $i$ and $t$ run over the same set of atomic environments.
To prevent overfitting, we use regularization. Since our data sets for the CHO materials have very few
entries (most notably our $GW$ data set), to collect error statistics
we test all of our models using $n$-fold
cross validation, where $n$ models are trained, each time leaving out one
of the data points, and the model is tested on the particular entry that is left out.
For the models based on QM9 data, for which many more training points are available,
we train ten different models for a given training set size, randomizing each time over
training configurations, and test on the remaining configurations.
We do not perform explicit hyperparameter optimization.

The basic ML model architecture used throughout this work is given by \eq{02}. The application
of this model is straightforward for learning the $GW$ and DFT predicted molecular C1$s$ and O1$s$ BEs
of our QM9 subset. For the latter, we have a large amount of $GW$ and DFT data, for which we can even
train models based on the $GW$ data alone. In addition, the data sets are ``coherent'' in the sense that
the DFT and $GW$ data sets are of equal size and that the computational data in both sets are well defined
for isolated structures.

However, model training and utilization become more
intricate for the CHO materials, where we have few $GW$ data, which are in addition only available
for the carved clusters. Our main objective for the CHO materials is to combine DFT and $GW$ data to i) improve
upon the accuracy of DFT and ii) overcome the current limitation of $GW$ calculations to
non-periodic systems. This implies that we must combine two or more datasets and potentially also two
or more ML models. We propose to compute a corrected binding energy (BE$^c$) for atom $i$ as
\begin{align}
\text{BE}_i^\text{c} = & \text{BE}_i^\text{DFT} \left(\{\textbf{r}_j^{\text{ext}}\}\right)
\nonumber \\
& + \left( \text{BE}_i^{GW} \left(\{\textbf{r}_j^{\text{carv}}\}\right) - \text{BE}_i^\text{DFT} \left(\{\textbf{r}_j^{\text{carv}}\}\right) \right),
\label{15}
\end{align}
where $\{\textbf{r}_j^{\text{ext}}\}$ denotes the atomic environment of $i$
within a periodic DFT calculation of the extended structures and $\{\textbf{r}_j^{\text{carv}}\}$ denotes a
\textit{truncated} representation of this environment, i.e., the one given by a carved
cluster centered on $i$. We have therefore split the input data in \eq{15} into two terms,
a baseline given by $b_i = \text{BE}_i^\text{DFT} (\{\textbf{r}_j^\text{ext}\})$, and
a correction given by $\Delta_i =  \text{BE}_i^{GW} (\{\textbf{r}_j^\text{carv}\}) - \text{BE}_i^\text{DFT}
(\{\textbf{r}_j^\text{carv}\})$.

The rationale for using \eq{15} is the following. First, the Fermi level alignment (important
in experimental solid-state XPS because the sample and detector are shorted) is provided by a neutral \dKS{}
calculation of the extended structure. In this neutral calculation the BE is computed for the transition
of an electron from the core level to the Fermi level; this is the type of \dKS{} calculation usually
performed for solid-state samples~\cite{ljungberg_2011,susi_2015,aarva_2019,aarva_2019b}. Second, the
\textit{correction} to this \dKS{} BE, that is, the difference between a $GW$ and a \dKS{} calculation
performed on \textit{exactly} the same system, is assumed to be i) local (justifying the use of carved
structures) and ii) independent of whether the core electron is excited to the vacuum level or the Fermi
level. While arguably intuitive, there is no formal reason, \textit{a priori}, why these assumptions
should hold true.
Instead, we verify their validity from the agreement between computational and experimental spectra reported
in Sec.~\ref{sec:results}.

With the data partition in \eq{15}, we
can choose two different routes for predicting BE$_i^\text{c}$, either i) train an ML model
from $\{b_i + \Delta_i\}$ or ii) train an ML model from $\{b_i\}$ and another
from $\{\Delta_i\}$, then obtain BE$_i^\text{c}$ as the sum of both predictions. It is not
straightforward to determine \textit{a priori} which option gives more accurate predictions, since
the learning rates and available amount of data points are different for each data set. We explore both
strategies for the CHO materials studied in this work. 
The outlined hybrid ML model architecture, i.e., combining data sets from both
periodic and cluster calculations, is further described in Section~\ref{sec:MLmodelsCHO}.

\section{Results and discussion}
\label{sec:results}
We start with a comparison of the $GW$ and DFT predicted excitations for the CHO molecules and
cluster models. Next, the performance of the ML models for the molecular excitations is discussed. We
proceed with results for the ML models, which are the building blocks for our hybrid ML architecture of
the CHO materials. We then demonstrate that the hybrid approach is key to achieve quantitatively
accurate XPS predictions of CHO materials for three showcases and introduce our XPS Prediction Server.

\begin{figure}
\centering
\includegraphics[width=\columnwidth]{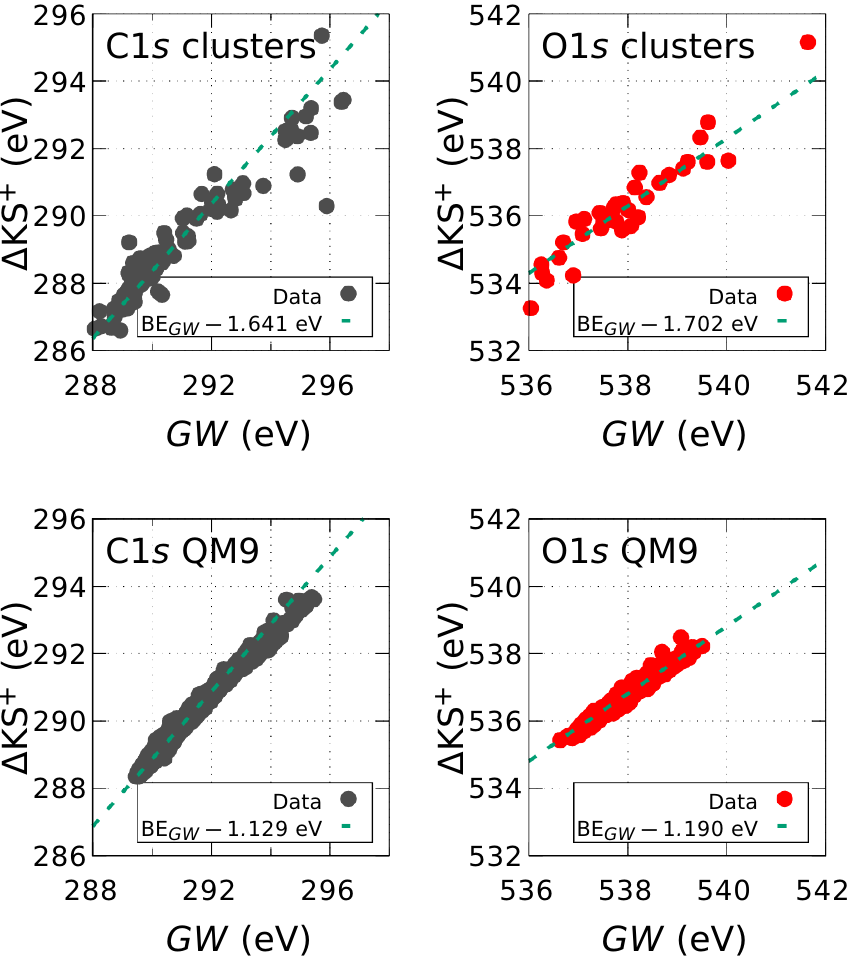}
\caption{Comparison between the $GW$ and \dKS{}$^+$ results obtained for carved clusters and
QM9 molecules. The dashed line indicates a linear fit, where the constant vertical shift
gives the leading difference between $GW$ and DFT data. This shift is specific
to each dataset and listed in the legend of each panel.}
\label{03}
\end{figure}
\subsection{Comparison of $\Delta$KS and $\boldsymbol{GW}$ excitations}
\label{sec:comparison}
The results of the $\Delta\text{KS}^{+}$ and $GW$ calculations are shown in \fig{03}, and compared to
each other, for both the cluster models of the solid-state CHO materials and our subset of small CHO-containing
molecules from the QM9 data set.
Figure~\ref{03} allows a direct comparison between the
$GW$ and DFT predictions, since they are computed on the exact same finite structures and are aligned both at the
respective vacuum levels. In all cases depicted in \fig{03}, the leading difference between $GW$
and DFT BEs, which we can identify as the leading error in the DFT prediction,
is a systematic underestimation of the BE. However, this leading error is dataset specific. It is in the
range of 1.1--1.2~eV for the QM9 subset and around 1.6--1.7~eV for the CHO clusters. In addition,
there are some subtle, but important, non-systematic differences.
In the remainder of this section we illustrate
how these subtleties and non-systematic differences in the data can be absorbed by our ML models, as
well as how these models can combine datasets to improve the
accuracy of the predictions.

\subsection{Learning molecular core-electron binding energies}
\label{sec:QM9learning}
\begin{figure}
\centering
\includegraphics[width=\columnwidth]{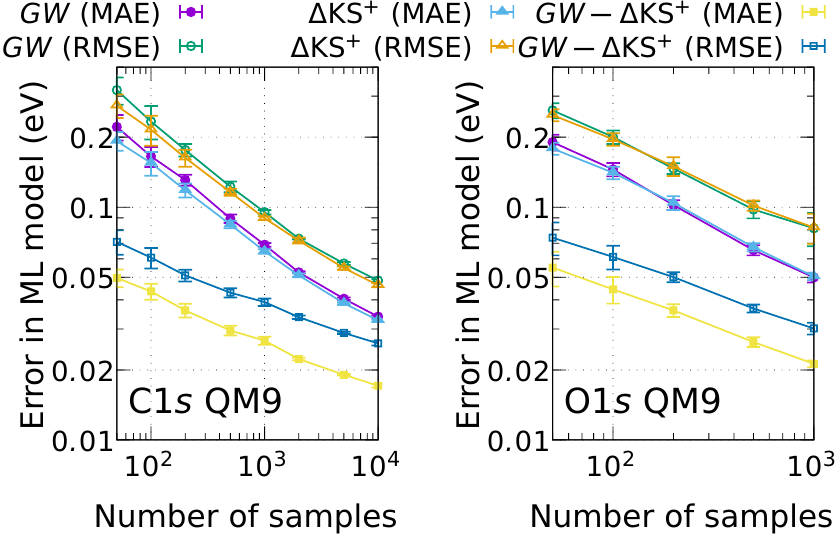}
\caption{Learning curves for different ML models based on BE data for CHO-containing
molecules in the QM9 database. $GW$ and $\Delta\text{KS}^{+}$ models show very similar learning rates,
and the $\Delta$ML model based on the difference between $GW$ and $\Delta\text{KS}^{+}$ demonstrates an extremely
good learning ability. For each training set size $n$ ten different models are trained,
each of which is constructed from $n$ randomly chosen training configurations.
The errors are then computed by testing the models on all the structures
not used for training, i.e., for a given model with $n$ training samples, the test set contains
$\text{14,707}-n$ and $1865-n$ samples for C and O, respectively. The error bars are computed by
averaging the errors over these ten different models.}
\label{04}
\end{figure}

\begin{table}[b]
\caption{Expected errors and timings when computing the core-electron BE of
a CHO-containing molecule with
four different approaches: i) a direct $GW$ calculation; ii) a {\dKS} calculation followed
by a correction based on a constant shift to account for the difference between $GW$ and {\dKS};
iii) a {\dKS} calculation followed
by a correction based on an ML model of the difference between $GW$ and {\dKS};
iv) the prediction of a ML model that learns the $GW$ result directly. The errors are (linearly)
extrapolated
from the learning curves in \fig{04} to the full size of our CHO-QM9 database: 14,707 and
1865 unique atomic environments for C and O, respectively. For simplicity, the $GW$ error
is taken as zero, and the other three approaches are designed to match the $GW$ prediction.
The CPU time refers to the average computational cost per molecule.}
\begin{ruledtabular}
\begin{tabular}{r | c | c c | c}
& $GW$ & \multicolumn{2}{c|}{\dKS$^+ +$} & $GW_\text{ML}$ \\
& & shift & $(GW-$\dKS$^+)_\text{ML}$ & \\
\hline
C1$s$ MAE (meV) & 0 & 75 & 15 & 27 \\
C1$s$ RMSE (meV) & 0 & 105 & 24 & 38 \\
\hline
O1$s$ MAE (meV) & 0 & 77 & 17 & 37 \\
O1$s$ RMSE (meV) & 0 & 95 & 25 & 61 \\
\hline
CPU time (s) & $\sim 300$k & \multicolumn{2}{c|}{$\sim 5$k} & < 1 \\
\end{tabular}
\end{ruledtabular}
\label{05}
\end{table}
In the following we demonstrate how to infer $GW$-quality core-level BEs from a DFT calculation based
on our QM9 subset of small CHO molecules.
Even though molecular excitations are not the target of this manuscript, the discussion of CHO molecules
is instructive because, even at the $GW$ level, these systems are small enough that plenty of data can be
generated and trends in ML accuracy and learning rates can be closely monitored.
We explore three different ways how to avoid an expensive $GW$ calculation, while retaining $GW$ accuracy.
(1) A $\Delta\text{KS}^+$ calculation is performed followed by a rigid shift of the obtained BE. (2) An ML
model for the difference between $GW$ and $\Delta\text{KS}^+$ results is developed and the ML predicted
difference is added to the result of the $\Delta\text{KS}^+$ calculation. (3) An ML model is trained that
learns the $GW$ data directly. The expected mean absolute errors (MAEs) and root mean square errors (RMSE)
with respect to the $GW$ reference are shown for all three approaches in Table~\ref{05} (the method for error
estimation is detailed below).

The first approach is motivated by the results in Section~\ref{sec:comparison}, where we found that the
leading difference between a $GW$ and $\Delta$KS prediction is a constant shift of the energies. If we take
a molecule from our QM9 subset and shift the $\Delta\text{KS}^+$ results by $+1.129$ (C$1s$) and $+1.190$~eV
(O$1s$), the prediction deviates on average by 75 and 77~meV from the $GW$ result, respectively; see also
Table~\ref{05} for the RMSE. These errors are already quite small. However, this approach still requires a
full \textit{ab initio} calculation at the DFT level, which is much cheaper than $GW$, but also becomes
computationally unfeasible for large disordered carbon structures.

We can improve the speed of the prediction and/or improve the accuracy of the prediction by using ML models. The learning curves
for $\text{BE}_i^{GW}$ and the difference $\Delta_i = \text{BE}_i^{GW}-\text{BE}_i^{\Delta\text{KS}^+}$ are
reported in \fig{04} for the C$1s$ and O$1s$ excitations of the CHO-QM9 subset. Displayed are the ML errors
(MAE and RSME) dependent on the number of data points used during training. The errors are computed by testing
the models on the portion of the entire database not used for training. For completeness, we included
also the learning curves for the \dKS{} computed BEs in Fig.~\ref{04}. The models for $\text{BE}_i^{GW}$
and $\text{BE}_i^{\Delta\text{KS}^+}$ exhibit similar learning rates with quickly decreasing errors.
The MAE is $< 40$~meV for both ML models when extrapolated to the limit of all the available data (training
plus testing). These errors are summarized in Table~\ref{05}.
Figure~\ref{04} also shows that learning the difference, $\Delta_i$, is easier than directly learning the
core-level BEs, as indicated by an extrapolated MAE of $<20$~meV. Moreover, the $\Delta$ML model achieves
the same accuracy with 50 data points as the $GW_{\text{ML}}$ model with 2000 data points. Finally, we note
that it is easier to learn the O1$s$ data than it is to learn C1$s$ data. This is due to the higher diversity
of possible C atomic environments.

Assuming the $GW$ value to be the ``golden'' standard, the most accurate prediction for CHO molecules
is obtained from the $\Delta \text{KS}^{+}+\Delta_i^{\text{ML}}$ approach, which reduces the
prediction errors to values in the vicinity of 20~meV. Such errors are negligible since they are an
order of magnitude smaller than the overall instrumental broadening in XPS experiments, and also smaller than
the smallest chemical shifts that can be resolved by analyzing experimental XPS spectra (i.e,
those that do not overlap).
The ML model trained from $GW$ data yields the next-best predictions. The worst approach is a
$\Delta \text{KS}^{+}$ calculation followed by a rigid shift, which has the same computational cost as
the $\Delta\text{ML}$-based scheme with an error that is $\sim 5$ times larger; see Table~\ref{05}.
Once the models are trained, the computationally cheapest prediction is obtained from the $GW_{\text{ML}}$
model, offering the best compromise between accuracy and speed. However, all three strategies are
computationally much cheaper than performing an actual $GW$ calculation,
which is already for small molecules almost two orders of magnitude more expensive than a $\Delta$KS
calculation (Table~\ref{05}). 

\subsection{Learning binding energies of CHO-containing materials}
\label{sec:MLmodelsCHO}
\begin{figure*}
\centering
\includegraphics[width=\textwidth]{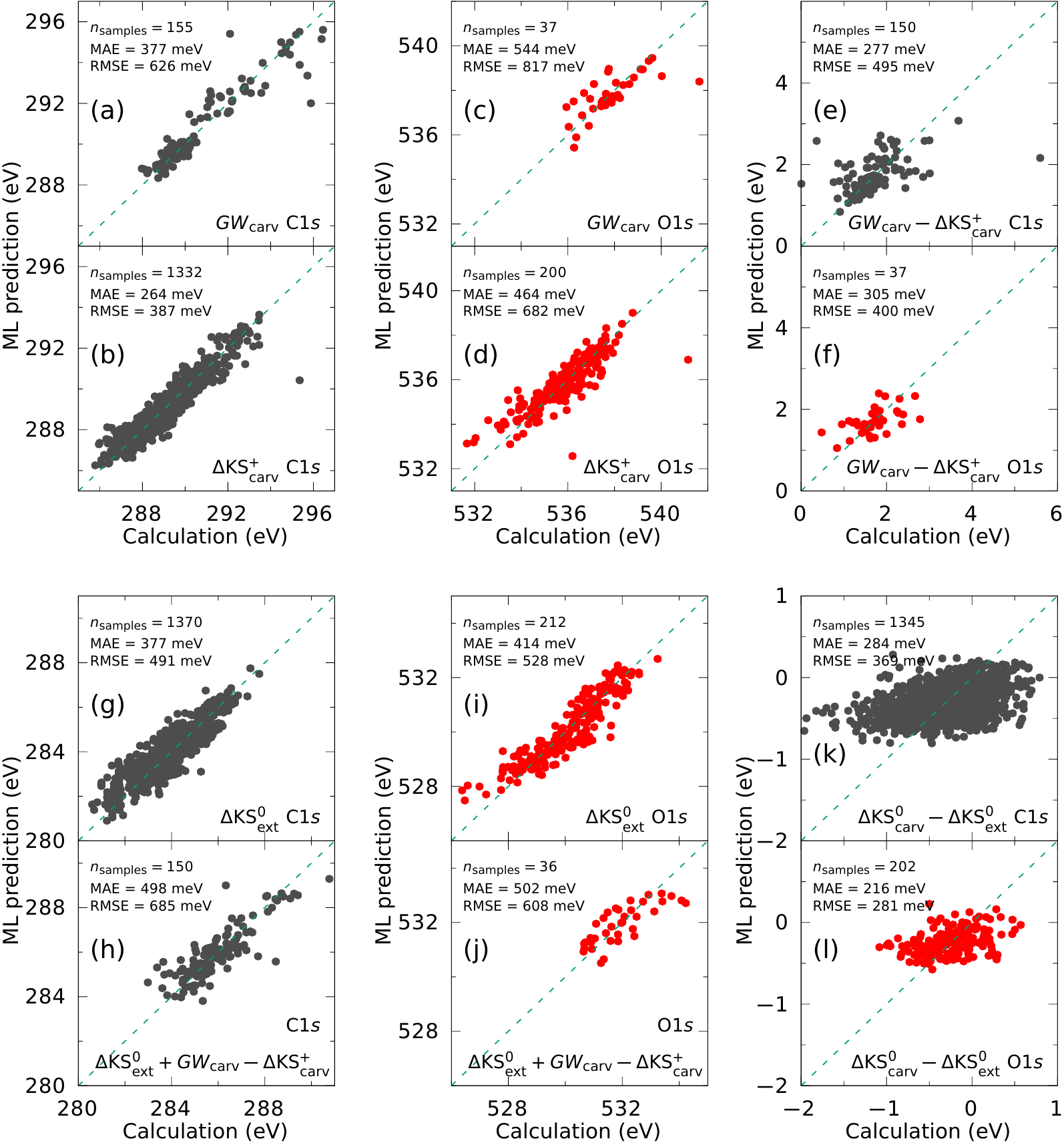}
\caption{Performance of the different ML models for C1$s$ (black) and O1$s$ (red) BEs trained as part of
this work. ``+'' and ``0'' refer to how the $\Delta$KS simulation is carried out in practice, i.e.,
by either removing the core electron from the sample or promoting it to the conduction band,
respectively. ``Ext'' stands for ``extended'' (periodic) structures, as opposed to carved
structures (``carv"). See text for a detailed discussion of the figure. The errors were obtained
by $n$-fold cross validation, due to the small size of the training sets.}
\label{06}
\end{figure*}

The generation of \dKS{} and, in particular, $GW$ data is significantly more expensive for CHO materials
than for molecules for the following reasons: i) The size of configuration space, i.e., the different ways
in which CHO atoms can be arranged in space, is much larger. This implies that more data are necessary to
train ML models of similar quality. ii) More atoms per site need to be considered to capture the effect of
the chemical environment on the excitation energy. This is true even when employing carved structure models,
where the number of atoms per atomic environment
can be still of the order of 100--200. If the scaling of the method of choice is $N$, this means
a cluster calculation is between $\sim 5^N$ and $10^N$ times more expensive than a QM9 molecule
calculation, since the cluster will contain 5--10 times more atoms than the largest molecules
in the QM9 data set.

Another problem is that the difference in the computational cost between DFT and $GW$ increases with
growing system size. In fact, the highest scaling steps in $GW$ only start to dominate the calculation for
structures larger than 30--50 atoms~\cite{Golze2018}, and hardly affect the computational cost for the CHO
molecules. For example, the computational time for a $GW$ calculation of a carved cluster with 96 atoms (out
of which 38 are C atoms) is 400,000 CPU hours, whereas the $\Delta$KS calculation takes approximately 22 CPU
hours for the same system. Compared to the molecular case (see Table~\ref{05}), the difference in computational
cost between $GW$ and DFT increased from a factor of 60 to 20,000.

For CHO materials, our \dKS{} and $GW$ databases are 1 and 2 orders of magnitude smaller,
respectively, than for the CHO-QM9 molecules.
However, the analysis of the CHO molecules in Section~\ref{sec:QM9learning} reveals the solution to this problem.
We have seen that a $\Delta$ML model, based on the difference between $GW$ and \dKS{} predictions, can be
trained to high accuracy with less data than directly training a model for the BEs at the \dKS{} or $GW$ level.
This justifies the strategy of developing hybrid ML architectures as outlined in Section~\ref{sec:MLmodel}.
Starting from Eq.~\eqref{15}, we consider two options: i) We learn a DFT baseline for the extended
(``ext'') structures and apply an ML-predicted correction based on the difference between $GW$ and DFT for the
carved (``carv'') structures as shown in Eq.~\eqref{08}. ii) We learn the $\Delta\text{KS}_{\text{ext}}^0$
baseline and the $\Delta$ term simultaneously as in Eq.~\eqref{10}:
\begin{align}
\left( \text{BE}^c \right)_\text{ML}  &\rightarrow
\left( \Delta \text{KS}_\text{ext}^0 \right)_\text{ML} +
\left( GW_\text{carv} - \Delta\text{KS}_\text{carv}^{+} \right)_\text{ML}
\label{08}
\\[5pt]
\left( \text{BE}^c \right)_\text{ML}  &\rightarrow
\left( \Delta \text{KS}_\text{ext}^0 +
GW_\text{carv} - \Delta\text{KS}_\text{carv}^+ \right)_\text{ML}.
\label{10}
\end{align}

The performance of the ML models required to construct the hybrid ML architectures in
Eqs.~\eqref{08} and ~\eqref{10} is shown in \fig{06}. We start with the discussion of the ingredients
for Eq.~\eqref{08}, i.e., the $\left( \Delta \text{KS}_\text{ext}^0 \right)_\text{ML}$ models for
the BEs of the extended CHO structures [\fig{06} (g) and (i)] and the $\Delta$ML models for the carved
structures [\fig{06} (e) and (f)]. For comparison, we trained also ML models for BEs of the carved structures
based on $GW$ [\fig{06} (a) and (c)] and $\Delta\text{KS}^+$ data [\fig{06} (b) and (d)]. 
We observe that the convergence of the $GW$ and $\Delta\text{KS}^+$ models for the BEs of the carved
structures is much slower than for the molecular case. For
instance, the best $\Delta\text{KS}^{+}_{\text{carv}}$ model for the C1$s$ excitations [\fig{06} (b)]
still shows a significantly larger error with over 1300 training samples ($\text{MAE} = 264$~meV), when
compared to the corresponding CHO-QM9 model in \fig{04} ($\text{MAE} \sim 65$~meV), corresponding to a four-fold
relative increase of the error. This is easily ascribed to the
much more complex configuration space spanned by CHO materials compared to small CHO
molecules, as discussed before. Nevertheless, we find that we can train $\Delta$ML models [\fig{06} (e)
and (f)] of reasonably good quality (MAE $\sim$ 300~meV) for the carved structures with as little as
150 (C$1s$) and 37 (O$1s$) data points. This is in line with the observation made for molecules that less
data are needed for the $\Delta$ML models. However, the leading error in Eq.~\eqref{08} will originate
from the $\left( \Delta \text{KS}_\text{ext}^0 \right)_\text{ML}$ model [\fig{06} (g) and (i)] with
a MAE of $\sim 400$~meV for both C$1s$ and O$1s$ BEs. Its learning behavior is in fact similar to the
$\left( \Delta \text{KS}_\text{carv}^+ \right)_\text{ML}$ model and the same arguments regarding the
complex configuration space apply.

The performance of the ML model where we train the DFT baseline and $\Delta$ term at the same time
[Eq.~\eqref{10}], is shown in Figs.~\ref{06} (h) and (j). Despite the small size of our training set
(150/36 data points for C$1s$/O$1s$), we obtain MAEs that are with 500~meV in the range of the
overall instrumental broadening in regular XPS experiments (synchrotron-based XPS experiments can
achieve better resolution).

The panels (k) and (l) in \fig{06} display $\Delta$ML models where we compare the core-level BEs
from {\dKS}$^0$ calculations of carved clusters to those of the corresponding periodic structures 
($\Delta\text{ML}=\Delta\text{KS}_{\text{carv}}^0 - \Delta\text{KS}_{\text{ext}}^0$). The results
in Fig.~\ref{01} (bottom) already indicated that the main difference between core-level BEs of carved and
periodic structures is a constant shift. The purpose of training these $\Delta$ML models is to assess
the validity of the locality assumption inherent to the carving process in more detail
(see Ref.~\cite{csanyi_2005} for a general discussion of locality
in atomistic modeling and Ref.~\cite{deringer_2017} for a discussion in the context of GAP
force fields). Compared to all other models displayed in \fig{06} (a-j), we find that the ML models
in panels (k) and (l) show the poorest \textit{relative} performance, i.e., the largest errors relative
to the spread of input values. The MAEs of 284~meV (C$1s$) and 216~meV  (O$1s$) are statistically
significant measures for the intrinsic errors due to the carving procedure. These MAEs quantify the
influence of the discarded portion of the periodic structure on the core-level BE. In other words, by
representing extended structures via
carved clusters truncated at 4.25~{\AA}, we will not be able to obtain predictions more
accurate than these errors, even in the limit of infinite data.
Fortunately, even though \textit{individual} errors from the carving procedure can be
expected in the order of 200~meV, the error in the statistical distribution of the predictions
is more significant for XPS prediction, since an XPS spectrum is constructed out of the
superposition of many individual BE contributions.
In addition, we never require $\Delta\text{KS}_{\text{carv}}$ to be an
accurate approximation of $\Delta\text{KS}_{\text{ext}}$. Instead, we need the difference
between $\Delta\text{KS}_{\text{carv}}$ and $GW_{\text{carv}}$ to be an accurate approximation
of the difference between $\Delta\text{KS}_{\text{ext}}$ and a hypothetical $GW_{\text{ext}}$
calculation, which we cannot carry out because periodic $GW$ core-electron BE calculations are
currently unavailable.

Taking the arguments for experimental broadening and
statistical distribution into account, we can indeed conclude that the clusters are reasonably good
surrogate models for the extended structures, a result that will
be corroborated in Section~\ref{12} for actual XPS spectra predictions. 

A final observation is that, while it may appear that it is easier to learn C1$s$ BEs than
O1$s$ BEs, this is solely due to the size of the training sets, which is in turn dictated
by the number of available C and O environments in the database. As we saw in
\fig{04} for the CHO-QM9 molecules, it is in fact easier to learn O1$s$ BEs. This is likely
due to the higher diversity of possible atomic motifs for carbon~\cite{caro_2018c} than for oxygen in
the CHO system.

\subsection{Predicting XPS spectra from the models}\label{12}

While linking molecular XPS spectra to the computationally predicted BEs from $GW$ and
$\Delta\text{KS}^{+}$ is straightforward, this connection is not so clear for materials.
There are two main differences that pose significant additional challenges.
The first difference is that, in experimental XPS of solid-state samples, the vacuum level
is not an easily accessible reference and the experimental BEs are typically reported with respect
to the Fermi level of the sample. In the context of electronic structure theory, the Fermi level is
only well defined for metallic systems. For semiconductors and insulators, we need to rely on the thermodynamic
definition, in which the Fermi level is given as the derivative of the total (free) energy with respect to
the number of electrons in the system.  As discussed in Section~\ref{13} and the SI, one possible
way to estimate the core-level BE with the Fermi level as reference is to perform a $\Delta\text{KS}^0$ calculation,
where we add the excited electron to the conduction band and relax the electronic structure. This is the strategy
we follow here and the reason why we learn a DFT baseline at the $\Delta\text{KS}^0$ level in Eqs.~\eqref{08}
and ~\eqref{10}.

The second difference arises precisely from the need for
a $\Delta\text{KS}^0$ calculation. The electron that we added to the conduction band will interact
with the core hole via the (screened) Coulomb potential leading to a spurious bound exciton. Compared
to valence band holes in semiconductors, the core hole is extremely localized and
the exciton BE will therefore be quite large (in the order of 0.5 to 1~eV for CHO materials)~\cite{gutierrez_1995}.
Since exciton binding stabilizes the system, the spurious exciton BE lowers the
{\dKS} prediction, compared to the actual core electron BE, i.e., the one that should be
compared to experimental XPS. Dynamical core-hole screening effects are also unaccounted for,
which can further complicate direct comparison with experiment. Fortunately, these exciton BEs tend
to be highly material-specific and lead to a constant shift of the whole computational XPS
spectra (towards lower values). Our future work will aim at quantitative estimation
of these exciton BEs for improved core-electron BE prediction.
Nevertheless, we find our models to be satisfactorily accurate, even in the absence of excitonic
corrections, for the purpose of comparing between computational and experimental spectra. We thus
speculate that the contribution of excitonic effects to the chemical shifts in
disordered carbon materials may be small enough to not affect this comparison.  

The XPS spectra of the CHO materials are computed as the superposition of the individual, experimentally
broadened, signals of each atom in a given atomic structure (or ``supercell''),
\begin{align}
\text{XPS}(E) = \sum_{i=1}^{N_\text{atoms}} \delta(E-E_i; \sigma), \quad E_i = \left( 
\text{BE}(i; S_i) \right)_\text{ML},
\label{07}
\end{align}
where $E$ is an energy in the spectrum. Each signal $E_i$ is given for atom $i$ by an ML model for
periodic (``extended'') structures as a function of its
atomic environment $S_i$. The latter is characterized via \texttt{soap\_turbo} many-body atomic descriptors. 
The smearing function $\delta(E-E_i; \sigma)$ is chosen to account for thermal and instrumental
broadening. An appropriate choice of broadening function is, e.g., a normalized Gaussian with width
$\sigma \approx 0.5$~eV~\cite{aarva_2019b}.

We will test three different models for $E_i$ in Eq.~\eqref{07}, all of which implicitly use the
Fermi level as reference. i) The $(\Delta\text{KS}_\text{ext})_{\text{ML}}$ model is employed to predict
the BEs followed by a rigid shift. The shift is obtained from \fig{03} (top): C$1s$ excitations are shifted
by 1.641~eV and O$1s$ excitations by 1.702~eV. ii) The hybrid ML model introduced in Eq.~\eqref{08} is used
to compute the $GW$-corrected $\text{BE}^c$. iii) Equation~\eqref{10} is employed to obtain an ML prediction
for $\text{BE}^c$. The leading physical assumption for approaches ii) and iii) is that the $GW$
correction to the charged excitation energies carries over to the neutral excitation case for extended
structures.
Which of the two $GW$-corrected ML models is optimal strongly depends on the amount of available
$GW$ data compared to \dKS{} data. Even though \eq{08} has two sources of error, the error in the first term can be
made very small with enough $\Delta \text{KS}_\text{ext}^0$ data. In \eq{10}, the amount of $\Delta \text{KS}_\text{ext}^0$ data that can be
used is limited to those structures for which $GW$ data is also available, thus limiting
the amount of training data that can be reused.

\begin{figure}
\centering
\includegraphics[width=\columnwidth]{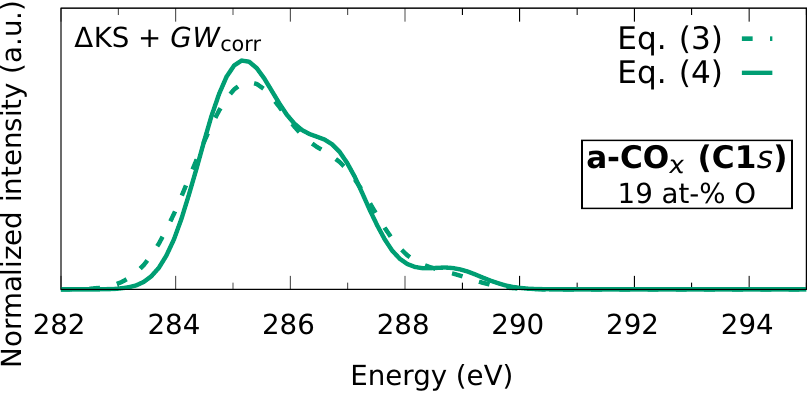}
\caption{Comparison of the $GW$-corrected ML models for the a-CO$_x$ with 19 at-\% O.}
\label{MLmodelcomparison}
\end{figure}

For the amount of training data
that we managed to gather for this work, both $GW$-corrected ML models  perform very similarly. We estimate the
RMSE for models based on Eqs.~(\ref{08}) and (\ref{10})
to be circa 0.697~eV and 0.685~eV, respectively, for C1$s$ predictions, where the error
for \eq{08} is estimated as the square root of the sum of the individual squared errors
(i.e., assuming the individual errors are normally distributed). For O1$s$, the estimated RMSEs
are 0.662~eV and 0.608~eV for Eqs.~(\ref{08}) and (\ref{10}), respectively. The similar performance
manifests also in the prediction of the XPS spectrum of the CHO materials. Figure~\ref{MLmodelcomparison} shows
that the predicted peak positions and overall spectrum shape are very similar. For the prediction of the XPS spectrum of
selected CHO materials in Section~\ref{sec:application}, we will use \eq{08} since this ML model has currently
more potential to be trained to even higher accuracy by gathering more $\Delta \text{KS}_\text{ext}^0$ data,
whereas a performance improvement with \eq{10} would require also additional $GW$ calculations.

\subsection{XPS spectra predictions for selected CHO materials}\label{14}
\label{sec:application}
\begin{figure*}
\centering
\includegraphics[width=\textwidth]{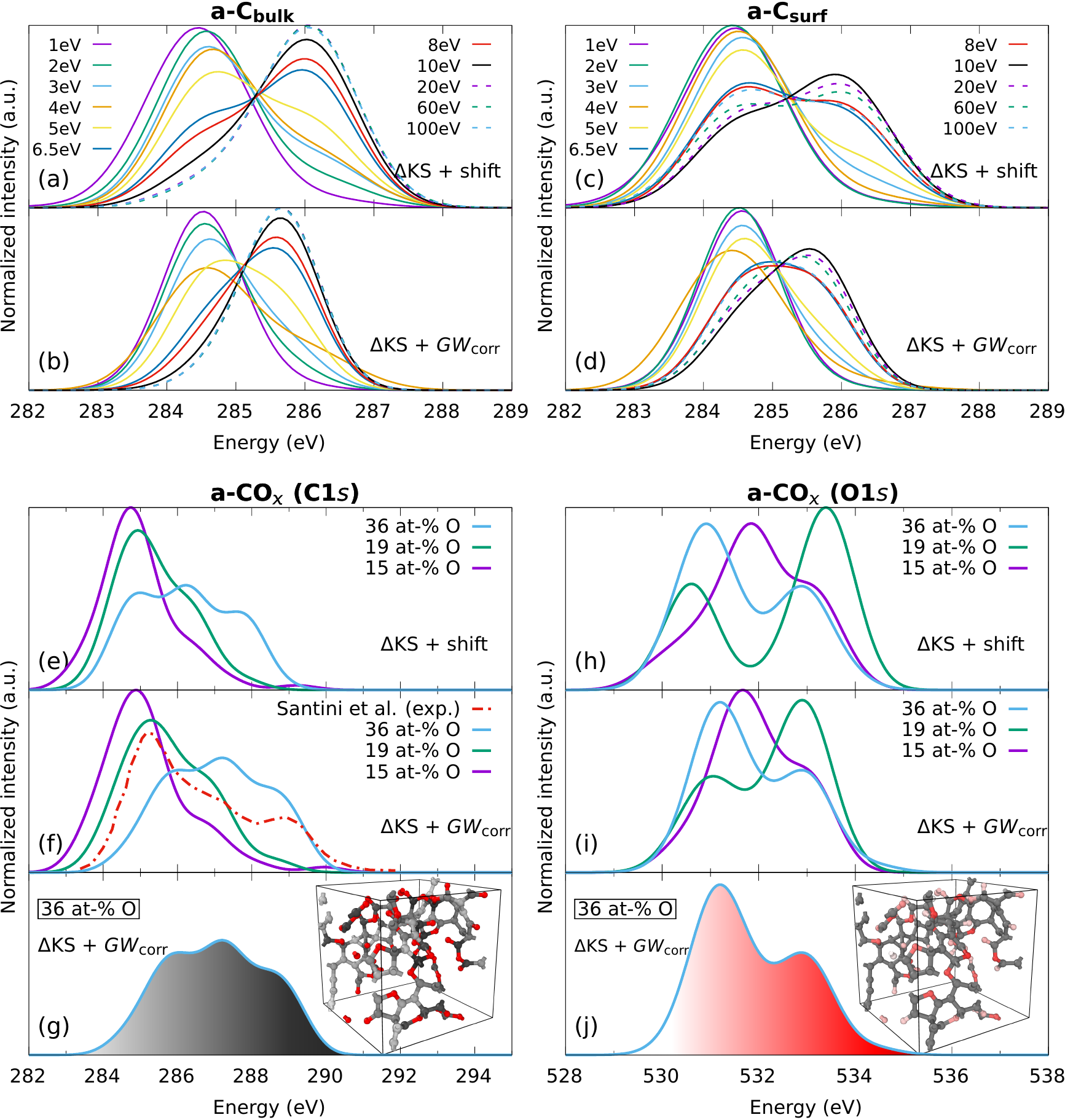}
\caption{XPS predictions based on our new methodology for C1$s$ spectra of (a,b) a-C bulk
and (c,d) a-C surfaces at different deposition energies and for (e-g) C1$s$ and (h-j)
O1$s$ spectra of a-CO$_x$. The panels (g) and (j) show site-resolved contributions to
the spectra. For example, the light-gray colored C atoms in panel (g) contribute to
light-gray regions in the spectrum, whereas dark-gray C atoms contribute to the dark-gray
spectral regions. The experimental a-CO$_x$ C1$s$ data was taken from Santini
\etal~\cite{santini_2015}. Compared are two models: $\left( \Delta \text{KS}_\text{ext}^0
\right)_\text{ML}$ predictions corrected by i) a constant shift (+1.641~eV for C$1s$ and +1.701~eV
for O$1s$) and ii) $GW$. All
$GW$-corrected predictions in this panel
are obtained with \eq{08}; see the SI for a comparison
of \eq{08} and \eq{10} for a-CO$_x$.}
\label{09}
\end{figure*}

The ultimate test for the models presented in this paper are predictions of XPS spectra
for realistic structural models of CHO materials and subsequent comparison to experiment.
We present XPS predictions for three classes of CHO materials: 1) a-C throughout the full range
of deposition energies, which in turn covers the full range of $sp^2$/$sp^3$ ratios
observed experimentally; 2) oxygenated amorphous carbon (a-CO$_x$) with different amounts of oxygen
content; and 3) rGO also with varying oxygen concentrations.

Experimentally, a-C thin films are grown by a number of physical
deposition methods~\cite{robertson_2002}, where the main deposition parameter is the kinetic
energy of the deposited atoms. Therefore, to model a-C realistically we use
computational structures generated in previous work for deposition energies in the range
1--100~eV~\cite{caro_2018,caro_2020c}.
The XPS predictions of the a-C structures are shown in \fig{09} (a-d), where we have focused
on two different regions of thin-film structures:
the bulk of the film [panels (a) and (b)], on the one hand, and the surface layer [panels (c) and (d)],
on the other. We present XPS predictions using \eq{07}
in combination with i) the $(\Delta\text{KS}_{\text{ext}}^0)_{\text{ML}}$+shift model and ii)
the ML model from \eq{08}. In the following, we refer to these models as $\Delta\text{KS}+\text{shift}$
and $\Delta\text{KS}+GW_{\text{corr}}$, respectively.

The deposition energies control the mass density and $sp^2/sp^3$ content of the a-C films. High
deposition energies yield films with high mass densities and $sp^3$ contents, whereas low deposition
energies correspond to low mass densities and high $sp^2$ content~\cite{robertson_2002,caro_2020c}.
This is also apparent from \fig{09} (b) and (d), where we observe a pronounced transition between an
$sp^2$-dominated XPS spectrum (peak at $\approx 284.5$~eV) for low deposition energy and an
$sp^3$-dominated (peak at $\approx 285.6$~eV) spectrum at higher deposition energy. The
turning point for the transition is between 5~eV and 6.5~eV incident atom energy. For the bulk, we
have a clear transition and the $sp^3$ peak at high deposition energies has the same intensity as
the $sp^2$ peak at low deposition energies. The transition is not fully developed for the surface layer,
where the intensity of the $sp^3$ peak is less pronounced. The reason is that the surface of the a-C
film generally contains lower coordinated atoms compared to the bulk. a-C films are $sp^2$-rich at the
surface even for very high densities and may contain significant numbers of undercoordinated $sp$ C
motifs, which are present only in negligible amounts in the bulk~\cite{caro_2018,caro_2020c}.

Comparing the $\Delta\text{KS}+\text{shift}$ and the $\Delta\text{KS}+GW_{\text{corr}}$ spectra in \fig{09} (a-d),
we find that the main effect of the $GW$ correction is to reduce the width of the predictions. The
$\Delta\text{KS}+\text{shift}$ model predicts the $sp^3$ peak to be
located 0.5~eV higher than the $GW$-corrected model.
Experimentally, the separation between $sp^2$ and $sp^3$ features
in a-C has been determined to be of the order of 1.1~eV~\cite{nagareddy_2018}. This is the same
separation predicted by our $GW$-corrected model, whereas the \dKS{} model predicts a separation
of circa 1.5~eV. The relative shifts from DFT-based $\Delta$-methods typically agree well with experiment
for small molecules~\cite{Bellafont2016}. However, this result indicates that the accuracy of the relative
shifts deteriorates for larger systems, which is a consequence of the delocalization error in DFT,
demonstrating the need for the $GW$ correction.

We discuss next the XPS predictions for a-CO$_x$ with different oxygen contents (15, 19 and 36 at-\% O).
Figure~\ref{09} shows the C$1s$ (e-g) and O$1s$ (h-j) excitations employing the $\Delta\text{KS}+\text{shift}$
and $\Delta\text{KS}+GW_{\text{corr}}$ model from \eq{08}. The main effect of the $GW$ correction is again the
reduction of the spread of the predictions,
especially for the O1$s$ spectrum. For a comparison between the $\Delta\text{KS}+GW_{\text{corr}}$ models from
of \eq{08} and \eq{10}, see \fig{MLmodelcomparison} and Fig.~S3 in the SI. The peak alignments
are very similar between the two $GW$ correction schemes, with slight differences regarding the relative intensity
and spread of the lower lying peaks.

The correspondence between excitation energy and
atomic motifs is highlighted in panels (g) and (j) of \fig{09} with color codings: light-gray (dark-gray) colored
C atoms in panel (g) contribute to light-gray (dark-gray) regions in the C$1s$ spectrum and light-red (dark-red)
colored O atoms in panel (j) to light-red (dark-red) regions in the O$1s$ spectrum. For C1$s$ spectra, the lower
energy contributions correspond to carbon-carbon bonds, followed
by an increase in the BE as the number of neighboring O atoms increases. The XPS spectra for a-CO$_x$ materials with
higher oxygen content have consequently more features at higher energies since the number of epoxide and ether (C-O-C),
keto (C=O) and ester (R-COO-R') groups increases. The core-level BEs of the C atoms in these groups increase also
in that order, where the largest C$1s$ excitation energies at around 289-290~eV are observed for carboxyl C atoms.
For the O1$s$ spectra,
the distribution is essentially bimodal. At lower energies we observe a peak corresponding to carbonyl O atoms from
keto or ester groups.
The peak at higher energies originates from contributions of the O atoms in epoxides and ethers
and the hydroxyl (singly-bound) O atom in the ester groups.
The relative intensity of these peaks strongly depends on the oxygen content. In our computational samples, epoxides,
ketos and esters are
present in approximate 13:61:26, 42:40:18, and 8:60:32
percentage ratios at 15, 19 and 36 at-\% O, respectively.

A comparison to experimental a-CO$_x$ data is available for the C$1s$ spectrum from Santini \etal~\cite{santini_2015}.
We observe
good agreement for the relative position of the different peaks present in the
C1$s$ spectra with our
$\Delta\text{KS}+GW_{\text{corr}}$ prediction. The agreement is less good for the peak intensities,
but the likely reason is that the relative concentrations of functional groups in the computational and
experimental samples are different.
We can infer from this direct comparison an oxygen content somewhere in between 19 and
36 at-\% (Santini \etal{} report $\approx 37$~\% for this sample),
and suggest that
a combination of those two simulated curves would lead to better agreement with experiment. This
in turn suggests that the experimental sample may be inhomogeneous with respect to the oxygen
content distribution. Reproducing the experimental structure more closely would
require deposition simulations similar to (but more complex than) those in
Refs.~\cite{caro_2018,caro_2020c}, which are non trivial and beyond the scope of this work.
In any case, determining the precise atomic percentages
experimentally is difficult because there are instrumental issues (such as
calibration), sample issues (heterogeneity, surface roughness), methodological issues (e.g.,
regarding how the peaks are fitted or how the background was subtracted) and many
more~\cite{shard_2020,gengenbach_2021,carvalho_2021}. Experimental
XPS-derived compositions will also often disagree with other methods, such as X-ray absorption
spectroscopy (XAS) or elastic-recoil detection analysis (ERDA), because of sample inhomogeneity
and different accessible depths. We show below for rGO that, when more candidate computational
structures are available, the atomic percentages can be resolved more precisely by matching predicted
and experimentally measured spectra. Therefore, the ability of ML-based XPS predictions to accurately
quantify atomic percentages in CHO materials may prove very useful in guiding and interpretation
of experiments.

\begin{figure}
\centering
\includegraphics[width=\columnwidth]{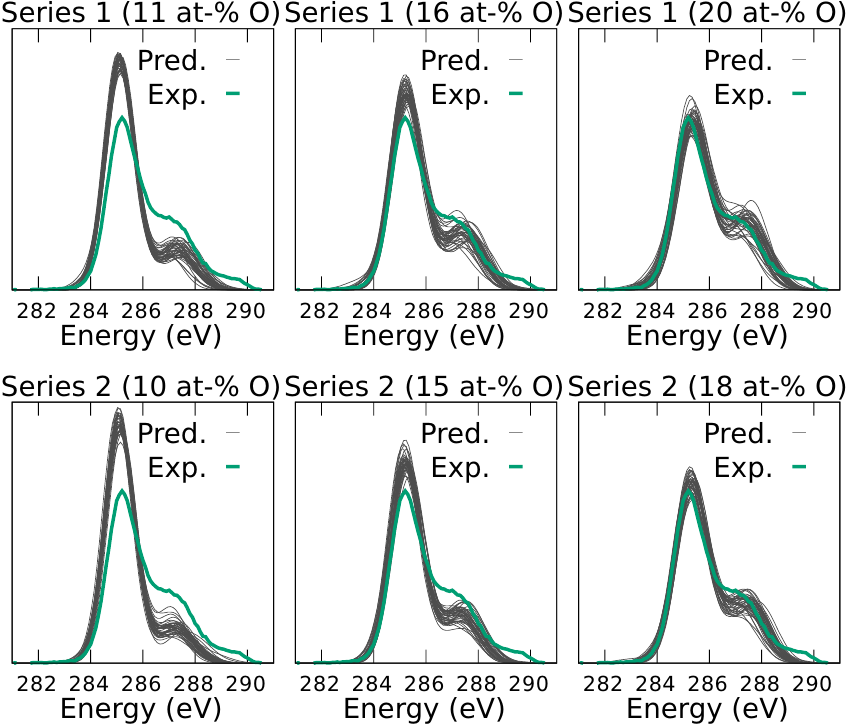}
\caption{Comparison of C1$s$ spectra between the computational predictions made with the present XPS
model [with $GW$ corrections, \eq{08}] for the rGO structural models of
Ref.~\cite{kumar_2013} and an experimental rGO XPS spectrum from Ref.~\cite{wester_2017}.
``Series 1'' and ``series 2'' indicates that the computational rGO models were
derived from different initial compositions of the GO precursor; series 1 was generated from
COOH-rich GO and series 2 from OH-rich GO~\cite{kumar_2013}. Note that the final content
of COOH in rGO for these models is negligible, regardless of the starting configuration used.}
\label{11}
\end{figure}

With our third application, rGO, we demonstrate how our developed methodology can be used to assess the
validity of candidate structural models for materials. The rGO structures in
our database were taken from Kumar \etal~\cite{kumar_2013} and contain different amounts of
oxygen in the range from 10 to 20~at-\%. They were either generated from COOH-rich GO (series 1) or
OH-rich GO (series 2) precursor structures. Altogether, there are 240 rGO structures with approximately
210 atoms each. We computed the XPS spectra of all of them using the $GW$-corrected ML model of \eq{08}.
Note that, altogether, these ML predictions take only a few minutes on  a desktop computer.

In \fig{11} we compare the predicted spectra for series~1 (top) and series~2 (bottom) to the experimental
rGO spectrum from Ref.~\cite{wester_2017}. From \fig{11} we can identify the candidate model structure
whose XPS spectrum best matches the experimental one. The spectra of candidate structures with a low oxygen
content of 10 or 11~at-\% clearly differ from experiment, while the ones with high-O content of 15--20~at-\%
agree best with the experimental XPS. Clearly, the experimental sample must contain a large fraction of oxygen. 
However, it is also evident that the structural models with high oxygen content are missing the functional groups
that contribute to the feature at $\sim 289.5$~eV in the experimental spectrum. As we saw for the
a-CO$_x$ example, this feature corresponds to carboxyl C atoms. These specific groups are not present in the rGO reference database
of Kumar \etal~\cite{kumar_2013},
even though the rGO structures from series~1 were generated from COOH-rich
GO starting configurations. However, unlike hydroxyl groups, the COOH groups are thermodynamically
unstable in the computational structure generation process and are not present in the final structural
models.
Our analysis thus indicates
that the composition of the experimental sample is similar, but not identical, to the
structural models with high O-content. In particular, it sheds light onto the missing bits of information -- in this case,
the presence of COOH groups.

\subsection{The XPS Prediction Server}

We have set up an online tool, which utilizes the different ML
models described throughout this paper. The XPS Prediction Server is available for
free at \href{http://nanocarbon.fi/xps}{nanocarbon.fi/xps}. The user can upload
a model structure in any format readable by the Atomic Simulation Environment
(ASE)~\cite{larsen_2017} and the
server will execute a Python script that runs the descriptor construction (via
calls to Quippy~\cite{libatoms}) and performs the kernel regression according to
a model of choice.
At the moment, only the CHO models described herein are available, but models for
other materials can be uploaded in the future as they are developed. The tool
works in a fully automated way, and for systems of usual sizes in the context of DFT modeling
of materials (a couple hundreds of atoms), a prediction can be obtained within seconds.
It is our hope to extend this concept of ML-based  computational prediction to other
materials and experimental observables, most notably other spectroscopic techniques.

\section{Summary and outlook}
\label{sec:summary}

We have presented an ML-based methodology to predict quantitatively accurate XPS spectra for
CHO containing molecules and materials. We generated a comprehensive database of computational
core-level BEs from DFT and $GW$ calculations. By careful combination of DFT and $GW$ data,
accurate ML models were trained for C$1s$ and O$1s$ excitations from relatively small data sets.
For molecular BEs, we showed that the errors in the ML predictions can be reduced to less than 50~meV.
The ML models were then applied to generate XPS spectra of selected CHO materials, namely a-C
thin films, a-CO$_x$ and rGO with different oxygen concentrations. Our predictions show excellent
qualitative and quantitative agreement
with experiment, resolving spectral shapes and features within 0.1~eV for the selected disordered
carbon-based materials. We also showed that ML
models trained with DFT data alone cannot reach this level of predictive power and that data
from the more accurate $GW$ approach are indeed crucial. For
disordered materials, we expect that whenever a suitably constructed database with  more $GW$ data is available,
an XPS ML model can be trained to provide accuracy close to
the practical resolution of common XPS experimental equipment.

We demonstrated the potential and suitability of our computational XPS tool to, e.g., quantify the atomic
percentages in a-CO$_x$ or identify shortcomings in candidate structure models of rGO. We have made this
new methodology freely available  to the public through the XPS Prediction Server. Such a computational tool may prove
valuable in guiding and interpreting experimental work, and in validating computational structural models of
materials. We hope to extend our ML models to other material classes
and spectroscopic techniques (XAS, Raman, IR, NMR, etc.) in the future.

\section*{Supporting Information}

The Supporting Information contains further details pertaining the calculations presented in this paper,
in particular technical information about the $GW$ and $\Delta$KS calculations. We include $GW$ convergence
studies for the C1s energies with respect to the cluster radius. We also show the MDS map of
CHO structures using a larger cutoff and a repeat the XPS spectra predictions using a different
ML model.

\begin{acknowledgments}
The authors acknowledge funding from the Academy of Finland under projects
316168 (D.~G.), 334532 (P.~R.), 310574, 329483 \& 330488 (M.~A.~C.), 321713
(M.~A.~C. and P.~H.-L.), and from their flagship program  Finnish Center for Artificial
Intelligence (FCAI), from the Emmy Noether Programme of the German Research Foundation under
project number 453275048 (D.~G.), from COST action CA18234, and from the European Research
Council (ERC) under the European Union's Horizon 2020 research and innovation
programme, under grant agreement no. 756277-ATMEN (T.~S.). Computing time from CSC
-- IT Center for Science, allocated for the Grand Challenge Project XPEC, is
gratefully acknowledged. Part of this work was carried out during a HPC-Europa3
mobility exchange (Horizon 2020 program under grant agreement 730897).
We thank V.~L. Deringer for providing the a-CO$_x$ structural models used in this study.
\end{acknowledgments}

\def\bibsection{}
\section*{References}

\begin{figure}
\centering
\includegraphics[]{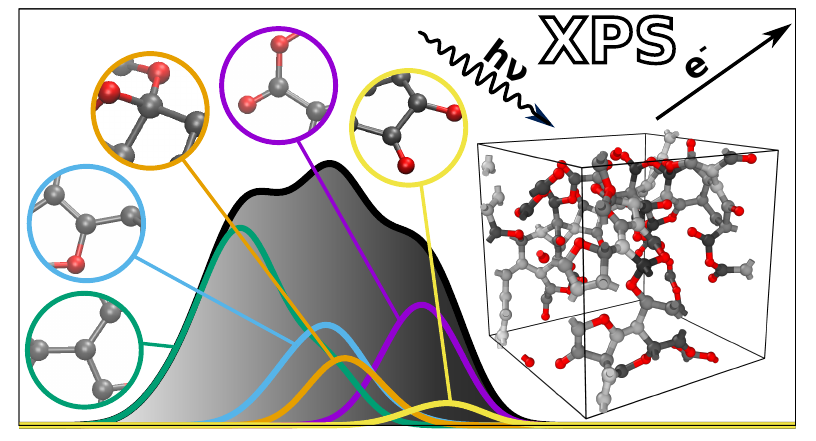}
\caption{ToC graphic.}
\end{figure}

\end{document}